\documentclass[twocolumn]{revtex4}
\usepackage{graphicx}
\usepackage{times}

\title{Design of a mode converter for efficient light-atom coupling
  in free space}

\begin{document}

\author{M. Sondermann}
\email{msondermann@optik.uni-erlangen.de}
\author{R. Maiwald, H. Konermann, 
  N. Lindlein, U. Peschel}
\author{G. Leuchs}
\email{leuchs@physik.uni-erlangen.de}
\affiliation{Institute of Optics, Information and Photonics, Max Planck
  Research Group, University of Erlangen-Nuremberg,\\
91058 Erlangen, Germany}

\begin{abstract}
In this article, we describe how to develop
 a mode converter that transforms a plane electromagnetic wave
into an inward moving dipole wave. 
The latter one is intended to bring a single atom or ion
from its ground state to its excited state by absorption of a single
photon wave packet with near-100\% efficiency.
\end{abstract}
\maketitle

\section{Motivation}
\label{sec:motivation}

The interaction of a single quantum of the electromagnetic field and a
single absorber is one of the fundamental processes in
optics and in physics in general.
Yet it is difficult to study the dynamics of an individual
absorption/emission process. In addition there is an apparent
asymmetry between emission and absorption. While an isolated excited
atom will definitely lead to the emission of a photon, the reverse
process is usually not very efficient: a single photon impinging on an
atom will only occasionally be absorbed, this corresponds to weak
coupling. 
Placing the atom in a cavity~\cite{raimond2001,walther2006}
constitutes one possibility of achieving strong coupling, and
there exist numerous experimental demonstrations (see, e.g.,
Refs.~\cite{goy1983,meschede1985,rempe1987,thompson1992,boozer2007}).
When attempting to achieve similar strong coupling in free space
without a cavity -- leaving the free space density of modes of the
electromagnetic field unmodified -- one could take guidance from the
emission process~\cite{quabis2000}.

In free space, the process of spontaneous emission of a single photon
is characterized by an exponential decay of the upper state population of the
emitter~\cite{weisskopf1930}, which is due to the interaction of 
the emitter with an infinite number of plane wave electromagnetic modes.
These modes can be considered as a heat bath~\cite{knight1997}, and
hence the process of spontaneous emission in free space is often
regarded as an irreversible process.
This irreversibility is, however, to be understood in a pure
thermodynamical sense, i.e., not as an in principle violation of time
reversal symmetry. The latter is expected only in connection with
CP-violation~\cite{lueders1957}.
The reversibility of the process can be inferred from the fact
that the Schr\"odinger equation is invariant under time reversal for a
closed system with a Hamiltonian without any explicit time dependence.
Since time reversal refers to an inversion of the
evolution in all degrees of freedom, we thus conjecture that a
single photon light field can be absorbed completely provided it was
generated by such an inversion~\cite{quabis2000}.

In the case of an atomic dipole transition, one should obtain this
effect -- absorption with a probability of one -- if one is able to
artificially create the time reversed version of the dipole wave
that is emitted by the atom, i.e., a dipole wave moving towards the
atom with its properties matched to the respective atomic dipole
transition.
Any deviation from the perfectly reverted wave reduces the achievable
probability for absorption as evidenced by calculations in
Ref.~\cite{alber1992}, where the temporal shape was not matched.
We can rephrase the underlying question as follows:
Is there a single-photon $\pi$-pulse that brings a system in
free space from its ground state to its excited state?

In Sec. \ref{sec:description} we describe how such a wave may be
created by use of a mode converter that basically consists of a
parabolic mirror and optical elements which tailor the spatial and
temporal distribution of the light field incident onto the parabolic
mirror.
The mathematical details of the procedure of obtaining the ideal
spatial field distribution have been discussed recently
elsewhere~\cite{lindlein2007}.
Therefore, the essential steps of this procedure are only reviewed
briefly here.
The mode converter proposed here may prove useful in quantum
computation schemes, where a reliable transfer of information from
photons to atoms or ions is desirable, in biophysical microscopy
applications, as well as in investigations of fundamental light-matter
interaction itself.
In Sec. \ref{sec:two-level} we discuss the influence of the energy
level structure of the absorber, highlighting the importance of a
clean two-level system for obtaining an absorption probability close
to unity and proposing possible atomic species which closely resemble
a two-level system.
Finally, in Sec. \ref{sec:comparison} the coupling scheme proposed
here is compared to already established coupling schemes.
 
\section{Description of the mode converter}
\label{sec:description}

\begin{figure}
\resizebox{0.45\textwidth}{!}{%
  \includegraphics{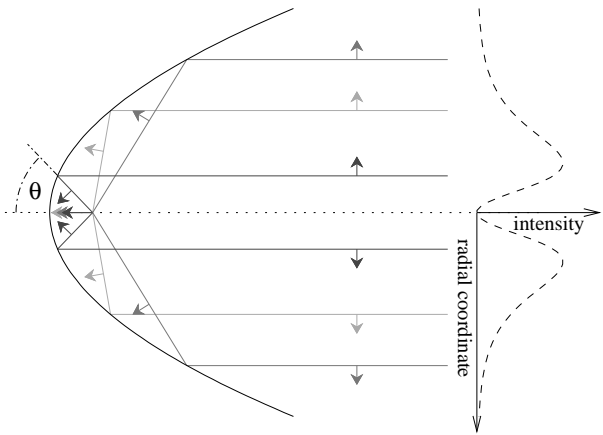}
}
\caption{Schematic drawing of the mode converter.
  Shown is a cut through the mirror (solid black line) along the
  optical axis.
  The dotted line denotes the optical axis.
  The gray lines are exemplary traces of incident light beams
  that travel parallel to the optical axis.
  The beams are then reflected by the mirror and meet at the mirror focus.
  The arrows denote the polarization vectors corresponding to each ray.
  The dashed line depicts the intensity distribution (arbitrary
  units) used for creating the radiation pattern of a linear dipole.
  The intensity is zero on the optical axis.
}
\label{fig:1}
\end{figure}

It has already been pointed out in
Refs.~\cite{quabis2000,vanenk2000,vanenk2004}, that the probability to
bring an atom from its ground state to its excited state upon
illumination strongly depends on the overlap of the illuminating light
wave with the dipole wave that corresponds to the atomic transition.
In the mode converter described here (see Fig.~\ref{fig:1}), this
overlap will be achieved by use of a deep parabolic
mirror~\cite{davidson2004}, as has been described
recently~\cite{lindlein2007}.
Due to the deepness of the mirror, the illumination of the absorber is
achieved from a solid angle of almost $4\pi$.

\begin{figure}
\resizebox{0.45\textwidth}{!}{%
  \includegraphics{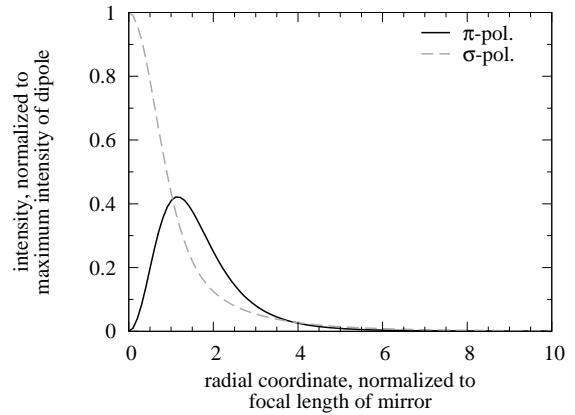}
}
\caption{Intensity distribution incident onto a parabolic mirror that
  creates the intensity pattern of a linear dipole ($\pi$-polarized
  light, solid line) or a circular dipole ($\sigma$-polarized
  light, dashed line) in the mirror focus.
  The quantization axis coincides with the optical axis of the mirror.
}
\label{fig:2}
\end{figure}

The emitter, e.g., an ion, is located at the focus of the parabolic
mirror.
To obtain the optimum illuminating intensity distribution, one starts
by studying the reverse process.
In case of a linear dipole oscillating parallel to the optical axis,
the emitted intensity pattern is proportional to
$\sin^2(\theta$)~\cite{jackson1999}, with $\theta$ being the angle
between the optical axis and the Poynting vector.
Each angle $\theta$ corresponds to a certain radial distance to the
optical axis after reflection off the mirror.
With this procedure one obtains the transverse intensity distribution
of the light field collimated parallel to the optical axis as a
function of the radial mirror coordinate (see Ref.~\cite{lindlein2007}
for a complete derivation). 
The resulting transverse distribution shown in Fig.~\ref{fig:2} is
precisely the input intensity distribution required for excitation.
This desired intensity distribution, which has considerable
overlap with a Laguerre-Gaussian mode of zeroth radial order and first
azimuthal order, can be obtained using a combination of two
diffractive optical elements~\cite{lindlein2007}.

In order to obtain the correct dipole wave in the mirror focus,
all partial waves have to arrive there with the same phase.
This is simply achieved by a flat phase distribution of the incident
wave in the case of a perfect para\-bolic mirror.
In reality, the parabolic mirror exhibits deviations from the
parabolic form leading to wave front aberrations.
Furthermore, the phase shift which an incident wave experiences upon
reflection is in general a function of the angle of incidence with
respect to the surface normal at a certain position on the mirror.
Both effects can be compensated for by placing an aberration
compensating phase plate in front of the mirror.
This ensures that after reflection off the mirror the incident wave
arrives at the focus with the same phase from all directions. 
Finally, an additional diffractive optical lens which transmits
most of the light in zeroth order and deflects some light directly to
the focus in the first diffraction order may compensate for the
otherwise incomplete angular distribution of the created dipole
pattern.
This has to be performed in such a way that the light directly
incident onto the atom and the light reflected by the mirror interfere
constructively.
However, even without such an element a parabolic mirror of a depth
of approximately six times the focal length already covers 94\% of the
light power required for the total radiation pattern of a linear dipole
oscillating along the mirror axis~\cite{lindlein2007}.

The polarization of the dipole pattern of the atomic transition under
investigation is created by tailoring the polarization of the incident
field.
For the case of a linear dipole oscillating along the optical axis of
the mirror, the incident field has to be radially polarized.
This is shown schematically in Fig.~\ref{fig:1}. 
In the case of a lens of high numerical aperture, radial polarization
was proven to produce a small focus with the focussed electric
field parallel to the optical axis in the focal
spot~\cite{quabis2000,dorn2003}.
Radial polarization can be generated e.g. by sending a linearly
polarized beam through a sectioned half wave
plate~\cite{ammon1996,quabis2005}, where the optical axis of the half
wave plate is rotated from segment to segment, or by use of liquid
crystal cells~\cite{quabis2005}. 

The inversion of an outgoing atomic dipole wave must be realized for
all relevant frequencies, i.e., over the entire atomic spectrum.
Thus, a single photon $\pi$-pulse that is meant to mimic such an inverted
wave has to have the spectral properties of the atomic transition. 
This is realized by a pulse of exponential shape, where the time
constant of the exponential function is the lifetime of the atomic
transition.
The carrier frequency of the pulse is the optical frequency of the
transition under consideration.
Since spontaneous emission manifests itself by an exponentially decreasing
probability to detect an emitted photon~\cite{mandel-wolf1995}, the
$\pi$-pulse that corresponds to the time inverted wave must have an
exponentially increasing shape. 

\section{Influence of energy level structure}
\label{sec:two-level}

The success of the coupling scheme employing a $4\pi$ mode converter
depends also on the
energy level structure of the emitter which is to be excited.
Let us assume there is more than one decay channel from the upper level,
say two as it is the case for an atom with $\Lambda$-like level
structure.
When exciting the system from one lower state, the system may decay
into the second lower state before completing the $\pi$-rotation of
the Bloch vector describing the transition from the first lower state
to the excited state.
Moreover, if there is more than one decay channel from the
excited state, the state of the emitter after the decay process is a
superposition of the ground states of all decay channels.
The corresponding emitted photon state is entangled with this atomic 
superposition state.
One possibility would be to wait for a spontaneously emitted
photon and to time reverse this single photon pulse and thus the whole
process by, e.g., phase conjugation.
Of course the time reversal of the atomic subsystem would pose a serious
difficulty.
Moreover, such a process would induce excess noise~\cite{gaeta1988}.
Therefore, we restrict the discussion to the ideal case of a simple
two-level system.

To obtain a true two-level system several criteria must be fulfilled.
First, there must be no hyperfine splitting, since otherwise the
ground state will be a multiplet and more than one decay channel
exist from the upper state.
This criterion calls for an atom with an even number of protons and
an isotope with an even number of neutrons.
Furthermore, the ground state must have a total angular momentum of
$J=0$ for the same reasons (this sets $J=1$ for the upper level).
Finally, the only dipole allowed transition from the excited
state must be the one to the ground state.
One possible candidate for a true two-level transition in neutral
atoms could be the $^1S_0\rightarrow^3\!P_1$ transition at 657~nm  in
neutral $^{40}$Ca.
If an ion is used it has to be doubly ionized.
Otherwise the requirement for a nucleus with an even number of protons
and neutrons and $J=0$ total electron angular momentum of the ground
state cannot be fulfilled. 
A possible candidate for such a transition is the
$^1S_0\rightarrow^3\!P_1$ transition at 252~nm in $^{174}$Yb$^{2+}$.

We note that similar strong coupling will also be achievable for a
$\Lambda$-like level structure when inducing the Raman coupling
between the two lower states with a bimodal spectrum of the exciting
laser fields as demonstrated in a cavity (see, e.g.,
Ref.~\cite{boozer2007}).
This coupling scheme~\cite{parkins1993} can also be transfered to
the free space mode converter situation. 

One could also think of conducting the experiment with an atom having more
than one ground state. This would require circularly polarized light to
optically pump the system into a cycling transition. But again the
circular polarization would lead to unwanted complications, such as an
angular emission-absorption pattern requiring light intensity on the
optical axis. 

\section{Comparison of the mode converter with other coupling schemes}
\label{sec:comparison}

A quantitative comparison between the photon-atom coupling in free
space and inside a resonator requires a properly defined figure of
merit.
In cavity quantum electrodynamics the established figure of merit is
$F_g=g^2/(\kappa\gamma)$ (see,
e.g., Refs.~\cite{miller2005,rempe1993}).
$F_g>1$ defines the strong coupling regime, where $g$, $\kappa$ and
$\gamma$ are the coupling between the atom and the cavity mode, the
cavity photon decay rate and the free space spontaneous emission
rate.
Miller et al. refer to $1/F_g$  as the 'number of strongly coupled
atoms necessary to affect appreciably the intracavity
field'~\cite{miller2005}.
This figure of merit is not immediately applicable to the free space
case, because $\kappa$ is not well defined in free space.
However, as can be deduced from the discussion in
Ref.~\cite{pinkse2002}, $F_g$ is largely determined by the geometry
of the cavity.
Adapting this train of thought, we compare in what follows the
photon-atom coupling achievable in free space to that demonstrated in
cavity quantum electrodynamics, which is related to the figure of
merit $F_g$.

In free space maximum coupling is achieved if the incoming photon wave
packet occupies the full solid angle which is $\Omega = 8\pi/3$,
including a weighting factor for the angular radiation pattern of the
linear or circular dipole transition.
Provided that the incident field has full overlap with the atomic
dipole radiation pattern~\cite{quabis2000,vanenk2004}, the coupling in
free space is maximum if the covered solid angle is equal to $\Omega$.
A reduced solid angle of $\Delta\Omega$ will result in an absorption
probability 
$P_{abs}=\Delta\Omega/\Omega=\frac{3\Delta\Omega}{8\pi}\le 1$.
Thus, the maximum absorption probability achievable with the proposed
mode converter is unity.

\begin{figure}
\resizebox{0.45\textwidth}{!}{%
  \includegraphics{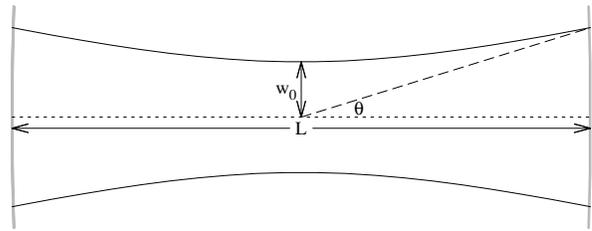}
}
\caption{Sketch of the geometry inside a resonator.
The gray solid lines depict the resonator mirrors.
$L$ is the resonator length.
$w_0$ is the beam waist of a Gaussian beam.
$\theta$ is the divergence angle of the beam.
}
\label{fig:3}
\end{figure}

We will now connect this approach to the cavity case.
The geometry of an atom residing inside a resonator is depicted in
Fig.~\ref{fig:3}. 
We assume that the atom is located at the position of the beam
waist and that it is illuminated by a single photon pulse of Gaussian
beam shape with beam waist $w_0$ and wavelength $\lambda$.
The divergence angle of the beam is $\theta$, which is taken to be
small with $1\gg\theta\approx\lambda/(\pi w_0)$.
Then, the solid angle from which the photon is incident onto the atom
is $\Omega_{cav}=\pi\theta^2$ for a single passage by the atom, where
we have implicitly supposed that the quantization axis of the atom is
oriented such that the angular radiation pattern has its maximum in
the direction of the cavity mirrors.
One thus arrives at a single passage absorption probability of
$\Delta\Omega_{cav}/\Omega=3\lambda^2/(8\pi^2w_0^2)$, which is a
number much smaller than one.
However, inside a cavity with a reflectivity $R$ of each mirror the
photon passes the atom $N=1/(1-R)$ times on average, which results in
a figure of merit 
\begin{equation}
\label{eq:Fomega}
F_\Omega=N\frac{\Delta\Omega_{cav}}{\Omega}=\frac{3\lambda^2}{8\pi^2w_0^2}
\cdot\frac{1}{1-R}\quad .
\end{equation}
For high quality factor resonators or high $R$, respectively,
$F_\Omega$ becomes much larger than one.
This reflects the situation commonly encountered in the strong
coupling regime of cavity quantum electrodynamics, where a photon is
absorbed and re-emitted repeatedly for many times before the photon
escapes the cavity.
One can thus define the probability that the photon is absorbed at
least once inside a cavity as 
\begin{equation}
\label{eq:Pcav}
P_{abs,cav}=\cases{ F_\Omega, &if $F_\Omega\le 1$\cr
1, &if $F_\Omega> 1$} \quad.
\end{equation}
This demonstrates that by employing the mode converter efficient
absorption should be possible in free space to the same extent as in a
cavity.

We note that it is straightforward to show that the two figures of
merit $F_g$ and $F_\Omega$ are identical.
Inserting the definitions of $g$~\cite{rempe1993},
$\gamma$~\cite{mandel-wolf1995}, using $\kappa\simeq
(1-R)c_0/L$ and approximating the cavity mode volume with $\pi w_0^2L$
proves the identity.

In the remainder of this section, we argue that the mode converter
does not modify the free space density of modes.
The presence of the surface of the parabolic mirror might be
expected to result in disturbing effects similar to the ones observed
in the case of emitters that are positioned close to surfaces within a
near field distance (see, e.g., Ref.~\cite{meschede1990} and citations
therein).
Since the distance of the atom/ion to the mirror surface in case of a
realistic parabolic mirror is much larger than a
wavelength, near field effects do not play a
role~\cite{lukosz1977,meschede1990}. 
In other words, the focussing mirror does not modify the 
density of modes at the focus.
This would be different if the atom was located on the optical axis at
a distance of the vertex radius of curvature away from the mirror
vertex -- a situation similar to the scenario examined experimentally
in Ref.~\cite{eschner2001} and theoretically in
Ref.~\cite{dorner2002}.
There, a finite fraction of the light emitted by an ion was refocussed
onto the ion by use of a lens and a mirror, leading to a change of the
spontaneous emission rate despite a large distance between ion 
and surface.
However, if the emitter is located at the mirror focus, the amount of
light which is back reflected from the mirror to the emitter
covers a vanishingly small part of the full solid angle and should
thus not have a considerable effect.
In the special case of a linear dipole oscillating along the
mirror axis, emission does not occur in this direction at all.
A modification of the spontaneous emission rate as observed for
emitters inside resonators~\cite{goy1983,kleppner1981} will therefore
not occur. 
Thus, the mode converter allows for investigations of light-atom
interaction in the 'natural' free space environment of the atom.

\section{Outlook}
There are several reports in the literature on experiments involving
optical excitation of single quantum systems in free space, including
far field excitation~\cite{wineland1987,vamivakas2007} as well as near
field excitation~\cite{gerhardt2007}.
The most successful free space coupling reported so far was
characterized by  12\% extinction of the excitation light by a
single quantum dot~\cite{vamivakas2007}.
However, in none of these experiments the quantum systems were
excited with single photon pulses tailored to the specific optical
transition in all aspects.
We believe that this will be possible with the mode converter
described here.
The use of this mode converter is expected to result in a coupling
efficiency or absorption probability, respectively, of almost 100\%.
The achieved absorption will sensitively depend on the
overlap of the generated single photon wave packet with a time reversed
dipole wave.

\begin{acknowledgements}
We acknowledge fruitful and stimulating discussions with 
G. Alber, J.C. Bergquist, J. Close, D. Meschede, S. Quabis,
L.L. Sanchez-Soto, W.P. Schleich, M. Stobinska, and D.J. Wineland.
\end{acknowledgements}

\end{document}